\documentclass[english]{IEEEtran}

\IEEEoverridecommandlockouts                              

\usepackage{subcaption}
\usepackage[utf8]{inputenc}
\PassOptionsToPackage{dvipsnames, table}{xcolor}
\usepackage{color}
\usepackage{array}
\usepackage{verbatim}
\usepackage{float}
\usepackage{multirow}
\usepackage{amsmath}
\usepackage{amssymb}
\usepackage{tabularx}
\usepackage{slashed}
\usepackage{svg}
\usepackage{ifsym}
\usepackage{wasysym}
\usepackage{tikz}
\usepackage{pgfplots}
\pgfplotsset{compat=1.17}
\usepackage{caption}
\usepackage{wrapfig}
\usepackage{diagbox}
\usepackage{booktabs}
\usepackage{cite}
\usepackage{url}
\usepackage{color}
\usepackage[dvipsnames]{xcolor}
\usepackage{tabularx}
\usepackage{color, colortbl}
\usepackage{soul}

\newenvironment{cmr}{\fontfamily{cmr}\selectfont}{\cmr}

\definecolor{LightCyan}{rgb}{0.88,1,1}
\definecolor{grey}{rgb}{0.95,0.95,0.95}

\renewcommand{\figurename}{\bf{Figure}}
\renewcommand{\tablename}{\bf{TABLE}}

\DeclareMathOperator{\diag}{diag}

\newtheorem{remark}{Remark}

\title{\LARGE \bf
Deep Neural Network (DNN) Based Nonlinear Adaptive Control of Quadrotors Operating Under Dynamic Uncertainties
}
\title{\LARGE \bf
Deep Nonlinear Adaptive Control for Unmanned Aerial Systems Operating under Dynamic Uncertainties
}

\author{Zachary Lamb$^{1}$, Zachary I. Bell$^{2}$, Matthew Longmire$^{3}$, Jared Paquet$^{3}$, Prashant Ganesh$^{4}$, Ricardo Sanfelice$^{1}$
\thanks{$^{1}$ Zachary Lamb and R. G. Sanfelice are with the Hybrid Systems Lab at UC Santa Cruz. Zachary Lamb was formerly with the Air Force Research Lab at Eglin AFB, FL 32542, USA.}
\thanks{$^{2}$ Zachary I. Bell is with the Munitions Directorate, Air Force Research Laboratory, Eglin AFB, FL 32542 USA. Email: \{zachary.bell.10\}@us.af.mil.}%
\thanks{$^{3}$ Matthew Longmire and Jared Paquet are with the Autonomous Vehicles Lab, University of Florida REEF, Shalimar, FL 32579. Email: {m.longmire@ufl, jared.paquet}.edu}
\thanks{$^{4}$ Prashant Ganesh is currently with EpiSys Science Inc. Formerly with the Autonomous Vehicles Lab, University of Florida REEF, Shalimar FL 32579.}
\thanks{This work is supported by NSF Grants no. CNS-2039054 and CNS-2111688, by AFOSR Grants no. FA9550-19-1-0169, FA9550-20-1-0238, and FA9550-23-1-0145, by AFRL Grant nos. FA8651-22-F-1052, FA8651-22-1-0017 and FA8651-23-1-0004, and by ARO Grant no. W911NF-20-1-0253.}
}

\begin{document}

\maketitle
\thispagestyle{empty} 
\pagestyle{empty}

\begin{abstract}

Recent literature in the field of machine learning (ML) control has shown promising theoretical results for a Deep Neural Network (DNN) based Nonlinear Adaptive Controller (DNAC) capable of achieving trajectory tracking for nonlinear systems. Expanding on this work, this paper applies DNAC to the Attitude Control System (ACS) of a quadrotor and shows improvement to attitude control performance under disturbed flying conditions where the model uncertainty is high. Moreover, these results are noteworthy for ML control because they were achieved with no prior training data and an arbitrary system dynamics initialization; simply put, the controller presented in this paper is practically modelless, yet yields the ability to force trajectory tracking for nonlinear systems while rejecting significant undesirable model disturbances learned through a DNN. The combination of ML techniques to learn a system's dynamics and the Lyapunov analysis required to provide stability guarantees leads to a controller with applications in safety-critical systems that may undergo uncertain model changes, as is the case for most aerial systems. Experimental findings are analyzed in the final section of this paper, and DNAC is shown to outperform the trajectory tracking capabilities of PID, MRAC, and the recently developed Deep Model Reference Adaptive Control (DMRAC) schemes.

\end{abstract}

\section{INTRODUCTION}\label{intro}


Linear controllers have drawbacks in controlling rigid body attitudes for Unmanned Aerial Systems (UAS's). For one, they are only effective around some small flight envelope near the equilibrium point, and two, they often require extensive lookup tables to control a flight over a wider regime \cite{fly-by-throttle} \cite{hinf} \cite{spaceGainSched}. Many nonlinear and adaptive control strategies exist to address these issues respectively \cite{nmpc}\cite{diffFlat}; however, fewer solutions exist which combine these two strategies into a single controller. Furthermore, some nonlinear control solutions require a prior accurate identification of various system parameters such as drag coefficients \cite{dragCoef}. These are often difficult to estimate and still do not address the issue of a changing model where these parameters are also uncertain. Seeing as how rigid bodies in many UAS applications are required to perform nonlinear maneuvers while undergoing stochastic model disturbances such as wind gusts, fuel slosh, or turbulence, there is a real need to build adaptive strategies into nonlinear controllers themselves.  

The objective of an adaptive controller is to improve tracking performance even when the dynamics of a system are uncertain or changing. By assuming a model and estimating of some set of parameters, control laws can be configured to change alongside the model. Model Reference Adaptive Control (MRAC) is one such adaptive scheme that works in a similar manner; by designing a reference model which one would want the real system to behave like, MRAC is able to optimize a set of control weights in order to produce system output which matches that of the reference model. This works by minimizing the loss between the reference model output and the true systems output. This technique is commonly applied in the field of aerospace. While MRAC works well, one key drawback is that it relies on a linear model assumption with few parameters.

Recently, a nonlinear adaptive Time Delay Estimation (TDE) based approach was employed to handle learning state-dependant uncertainties for a quadrotor \cite{Swati}. This method reduced the necessity for a priori knowledge of the system and enabled the learning of more natural nonlinear dynamics through a set of parameters. This method has shown promising results, and its implementation requires only a small amount of intuition on the bounds of model parameters which are specific to each dynamic system. However, applying modern neural-network based learning strategies would reduce the required model information even further, facilitating simpler generalization between systems, while also potentially allowing for more complex estimation of the uncertainties in the dynamics. To this effect, neural networks have been used in recent times to replace traditional model-based learning strategies for their empirical improvement over other system identification methods \cite{Ian}.


Neural networks which include one or more hidden layers are defined as Deep Neural Networks (DNN), and are often regarded as Universal Function Approximators (UFA) for their ability to approximate any function. However, while neural networks are powerful, many of their controller based applications lack performance guarantees during the training phase. Proximal Policy Optimization (PPO) is one such example of a Reinforcement Learning (RL) algorithm that lacks these properties. PPO typically requires many thousands or even  millions of training cycles to become effective at the task being trained on. During this training phase the target system will behave undesirably, and in the case of training an attitude controller onboard a quadrotor, this leads to dangerous crashes. Many RL control techniques do exist for quadrotors; however, these control strategies tend to be implemented at a high level for path planning because of the aforementioned reasons \cite{racing} \cite{quadRL}. Recent research into lower level control investigated a Deep RL (DRL) PPO attitude control technique which was able to achieve ``satisfactory'' performance in 9 hours of simulation training \cite{rlAttitude}. Even after \cite{deepTrain} and \cite{cascade} improved the training time to only 50 and 17 minutes respectively, this training phase is still too long if the end goal is to apply the technique to hardware. A quadrotor needs the ability to learn new flight environments in real-time to prevent crashes; unless a high enough fidelity model is incorporated to simulate the system during the training phase, this necessarily means that ML techniques such as DRL and PPO are difficult to implement at low levels in safety-critical systems such as a quadrotor's Attitude Control System (ACS).

Building on these overarching issues of a slow time-to-learn and lack of stability guarantees in some machine learning control policies, \cite{Chowdhary} designed a DNN based Model Reference Adaptive Controller (MRAC) which was coined ``Deep MRAC'' (DMRAC). By marrying the ability of a DNN to learn models with the the ability of MRAC to provide bounded stability guarantees, a stable adaptive controller was designed to properly allow DNN based control to thrive on a safety-critical system at low-levels of the feedback loop.

However, as was a drawback with traditional MRAC, DMRAC forces linear model tracking behavior. While tracking is obviously good, linear model based controllers are most effective at tracking trajectories near the point of linearization. For UAS's required to perform nonlinear attitude maneuvers, a nonlinear control strategy is preferred. In an attempt to improve on DMRAC, \cite{Bell2} designed DNAC, a generalized nonlinear adaptive controller which uses a DNN to estimate the model uncertainty. This design addresses issues associated with linear controllers, and shows promising trajectory tracking results for nonlinear systems in simulation. In DNAC, the Lyapunov analysis combined with a DNN provides some performance guarantee's that an RL based controller may not have.

Building off of these trajectory tracking results, this paper features an easily configurable control architecture for UAS's which significantly improves tracking performance and can adapt to changes in the model. The contributions of this paper are as follows:

\begin{enumerate}
    \item DNAC's setup is generalized and practically modelless, allowing it to be applied with little to no changes to any system.

    \item The unique multiple-timescale learning structure of the DNN allows for fast real time updates to the model, resulting in a controller capable of tracking desired trajectories under dynamic uncertainties with no prior training and trivial system knowledge.

    \item Most applications of ML based controllers lack boundedness guarantees. This approach benefits from the important learning capabilities of a DNN, while maintaining stability guarantees necessary to control a safety critical system.
\end{enumerate}


\newpage
This paper begins with necessary background in Section \ref{problemSetup}, providing adaptive update and control law algorithms from \cite{Bell2} as well as describing aspects of the DNN design. Section \ref{controlPlatform} dives into the quadrotor control platform, including details about the onboard hardware and software. Finally, Section \ref{results} begins with a description of DNAC's configuration, and then presents and compares real-world hardware results for PID, MRAC, DMRAC, and DNAC attitude controllers. These controllers are run for experiments where the quadrotor attempts to track a desired trajectory whilst undergoing heavy model disturbances which include a crosswind, wall effect, added mass, and sloshing effect.

\section{BACKGROUND}\label{problemSetup}
\subsection{System Dynamics}
In this paper, nonlinear systems with dynamics affine in control input are considered given by
\begin{eqnarray}
\dot{x} &=& f(x) + g(x)u,\label{dyn}
\end{eqnarray} 
where $x\in \mathbb{R}^n$ is the state of the system with $n \in \mathbb{Z}$ components, $g : \mathbb{R}^n \rightarrow \mathbb{R}^{n\times n}$ is the control effectiveness function, $u\in \mathbb{R}^n$ is the control input, and  $f : \mathbb{R}^n \rightarrow \mathbb{R}^n$ represents the drift dynamics which are usually uncertain.  Often times the underlying dynamics do not reflect the model exactly; however, in practice, the control affine structure is sufficient to approximate the model. Additionally, the true system model is typically unknown and assumptions are made enabling the utilization of adaptive control methods. The recent result in \cite{Bell2} proposes a method to estimate $f$ using a DNN and demonstrates how to estimate $g$ using a novel control structure. Expanding this concept to systems where the control effectiveness is constant and full rank, it follows that a constant $g$ can be utilized as a user-defined, diagonal gain matrix. The DNN compensates for the uncertainty of the system enabling the DNN control structure to add stable adaptability.

Motivated by \cite{Bell2}, the model uncertainty $f$ is modeled via a DNN as
\begin{eqnarray}
f(x) &=& W^\top\sigma(\Phi(x)) + \epsilon(x),
\end{eqnarray}
where, by the Universal Function Approximation (UFA) property \cite{UFA}, $W \in \mathbb{R}^{L \times n}$ is the unknown ideal output layer weights, $L \in \mathbb{Z}$ denotes the user-defined number of neurons in the output layer, $\sigma: \mathbb{R}^p \xrightarrow{} \mathbb{R}^L$ is the output layer activation function, $\Phi: \mathbb{R}^n \xrightarrow{} \mathbb{R}^p$ is the ideal unknown DNN features function, $p \in \mathbb{Z}$ denotes the user-defined number of inner layer features, and $\epsilon: \mathbb{R}^n \xrightarrow{} \mathbb{R}^n$ is the bounded unknown function reconstruction error. The ideal unknown DNN features function is expressed as a function composition of the inner layer weights and activation functions, namely, $\Phi(x)=(\phi_k \circ \phi_{k-1} \circ ... \circ \phi_1)(x)$, where $\phi_l=\sigma_l(W^\top_l\phi_{l-1}+b_l),\hspace{3pt}l\in \{1,2,...,q\}$, $q\in\mathbb{Z}$ denotes the number of inner layers of the DNN, $W_l\in\mathbb{R}^{L_{l-1} \times L_l}$ represents the $l^\text{th}$ inner layer weights with $L_l$ neurons, and $\phi_l$ denotes the $l^\text{th}$ inner layer activation function. This yields the following approximation of the model uncertainty:
\begin{eqnarray}
\widehat{f}(t,x) &\triangleq& \widehat{W}^{\top}(t)\sigma(\widehat{\Phi}(x)),\label{fHat}
\end{eqnarray}
where $\widehat{W}(t)$ and $\widehat{\Phi}(x)$ are the current estimates of $W$ and $\Phi(x)$, respectively. Updates to $\widehat{W}(t)$ and $\widehat{\Phi}(x)$ in (\ref{fHat}) is subsequently broken into two stages. The output weights $W(t)$ are estimated online in real-time using a Lyapunov-based adaptive update law. The inner layer weights, $\Phi(x)$, are estimated using batches of data collected online, for which a user-defined loss function is optimized using the extended stochastic gradient decent algorithm Adam \cite{Adam}. The combination of online and batch updates allows for fast, real-time adaptation of the outer-layer weights of the DNN, and slower-timescale learning for inner-layer weights to learn the latent features of the nonlinear dynamics. 

While the DNN compensates for the model uncertainty term $f(x)$ from (\ref{dyn}), the user-defined control effectiveness matrix is selected as
\begin{eqnarray}
\widehat{g}(x) &\triangleq& \diag(\widehat{g}_1,\widehat{g}_2,...\widehat{g}_n),\label{gHat}
\end{eqnarray}
where $\widehat{g}_i, \hspace{3pt} i\in \{1,...,n\}$ is the $i$th element of the control effectiveness gain matrix. Since this matrix is inversely proportional to the input in the subsequently designed control law, $\widehat{g}$ can be designed as another tunable gain matrix to adjust control effort.
\begin{remark}
Experimental results in Section \ref{results} will demonstrate that utilizing the assumptions made for $\widehat{f}(x)$ and $\widehat{g}(x)$ is sufficient to derive a controller which adeptly learns to stabilize a system with unknown and or changing dynamics. However, if desired, a more accurate $\widehat{g}(x)u$ term can be configured to adaptively update using system ID techniques following Assumption 1 in \cite{Bell2}.
\end{remark}

\newcommand\mymapsto{\mathrel{\ooalign{$\rightarrow$\cr%
  \kern-.15ex\raise.275ex\hbox{\scalebox{1}[0.522]{$\mid$}}\cr}}}
  
\subsection{Control Design}
The control objective is to minimize the norm of a time-varying trajectory tracking error for a given reference $t \mymapsto x_d(t)$. For a state trajectory $t \mymapsto x(t)$, the tracking error is defined as
\begin{eqnarray}
e(t) = x_d(t) - x(t),
\end{eqnarray}
where $e : \mathbb{R}_{\geq 0}\rightarrow \mathbb{R}^n$. To achieve this goal, the control design determines a robust and adaptive control law that utilizes the DNN estimates of the model uncertainty and guarantees trajectory tracking for (\ref{dyn}), namely, $e \rightarrow 0$ as $t \rightarrow \infty$. The Lyapunov analysis for the subsequent control and update law follow \cite{Bell2} closely and are therefore left out of this paper; however, the resulting control law $u(e,x,t)$ and outer-layer weight update law $\dot{\widehat{W}}(e,x,t)$ were derived to drive the tracking error to zero.

\begin{eqnarray}
u(e,x,t) \triangleq \widehat{g}^{-1}(-Ke - K_s\text{sgn}(e) + \dot{x}_d(t) - \widehat{f}(x,t)),\label{controlLaw1} \\ 
\dot{\widehat{W}}(e,x,t) \triangleq \Gamma_W\sigma(\widehat{\Phi}(x(t)))e^\top(t), \label{controlLaw2}
\end{eqnarray}
where $K$, $K_s$, and $\Gamma_W$ are user-defined tuning parameters. 

\begin{remark}
Generally, a large $\Gamma_W$ is associated with the ability to quickly learn the system dynamics, and a large $K$ is associated with fast trajectory tracking convergence. In practice, if $\Gamma_W$ is too small, then convergence of the model uncertainty term may be slow. As a result, a large $K$ will amplify undesirable signals present in $\widehat{f}$. For this reason, one should first tune $\Gamma_W$ for state estimation convergence, then tune $K$ for tracking performance. $K_s$ is a sliding mode control and should remain positive but close to 0. User-defined training parameters for updating the inner-layer weights of the DNN, such as the number of epochs, activation functions, and number of hidden layers, are  selected to enable the DNN to learn the nonlinear and slower timescale components of the dynamics.
\end{remark}

\subsection{DNN Inner-Layer Design and Update}
The inner-layer weights are updated in batches after a replay buffer has reached the set memory size. To perform the $j^{\text{th}}$ batch update to the inner-layers, data samples are stored in the $j^{\text{th}}$ replay buffer, $\mathcal{B}_j \triangleq \{B_k\}_{k=1}^{M}$, where $B_k=[\dot{x}_k, x_k, g(x_k)u_k]$, and $\dot{x}_k, x_k, u_k$ represent $\dot{x}(t_k), x(t_k), u(t_k)$, $t_k \in [t_{j-1}, t_j]$, $k \in [1,2,...,M]$, $M$ is the user-defined memory size, and $t_{j-1}$ is the time the last update to the inner-layer features completed. Once the $j^{\text{th}}$ buffer has stored $M$ samples, an update is performed for the inner-layer weights using the $j^\text{th}$ batch of data in $\mathcal{B}_j$ to compute the Smooth L1 loss
\begin{equation}
  \mathcal{L}_j=\frac{1}{M}\sum_{k=1}^M\mathrm{L}_k,\label{totalLoss}
\end{equation}
where
\begin{equation}
  \mathrm{L}_k =
    \begin{cases}
      0.5 \frac{(\dot{\widehat{x}}_k - \dot{x}_k)^2}{\beta}, & \text{if $|\dot{\widehat{x}}_k - \dot{x}_k|$ $< \beta$},\\
      |\dot{\widehat{x}}_k - \dot{x}_k| - 0.5\beta, & \text{otherwise},\label{lossFunc} 
    \end{cases}       
\end{equation}
\begin{eqnarray}
\dot{\widehat{x}}_k=\widehat{f}(x_k)+\widehat{g}(x_k)u_k\label{xPred}
\end{eqnarray}
Here, $\dot{\widehat{x}}_k$ is a DNN-based estimate of $\dot{x}_k$, which along with the smoothing parameter $\beta$, is used in calculating the loss via (\ref{totalLoss} and \ref{lossFunc}). After computing the loss, an optimization step is taken to update the inner-layer weights via Adam.

\section{QUADROTOR CONTROL PLATFORM}\label{controlPlatform}
The problem setup in Section \ref{problemSetup} is generalized for controllable states of nonlinear systems to track time-varying trajectories, for example, the attitude and angular velocities of a quadrotor. To demonstrate the performance of this control design, quadrotor flight experiments are performed for which a variety of uncertain model disturbances are present. An overview of the quadrotor control setup and position-level feedback loop will be described in Subsection  \ref{ss:system architecture}. The Robot Operating System (ROS) is used for message passing and facilitates this entire feedback structure.

\begin{figure*}[h]
    \centering  
    \includegraphics[height=2.5in, width=5.2in]{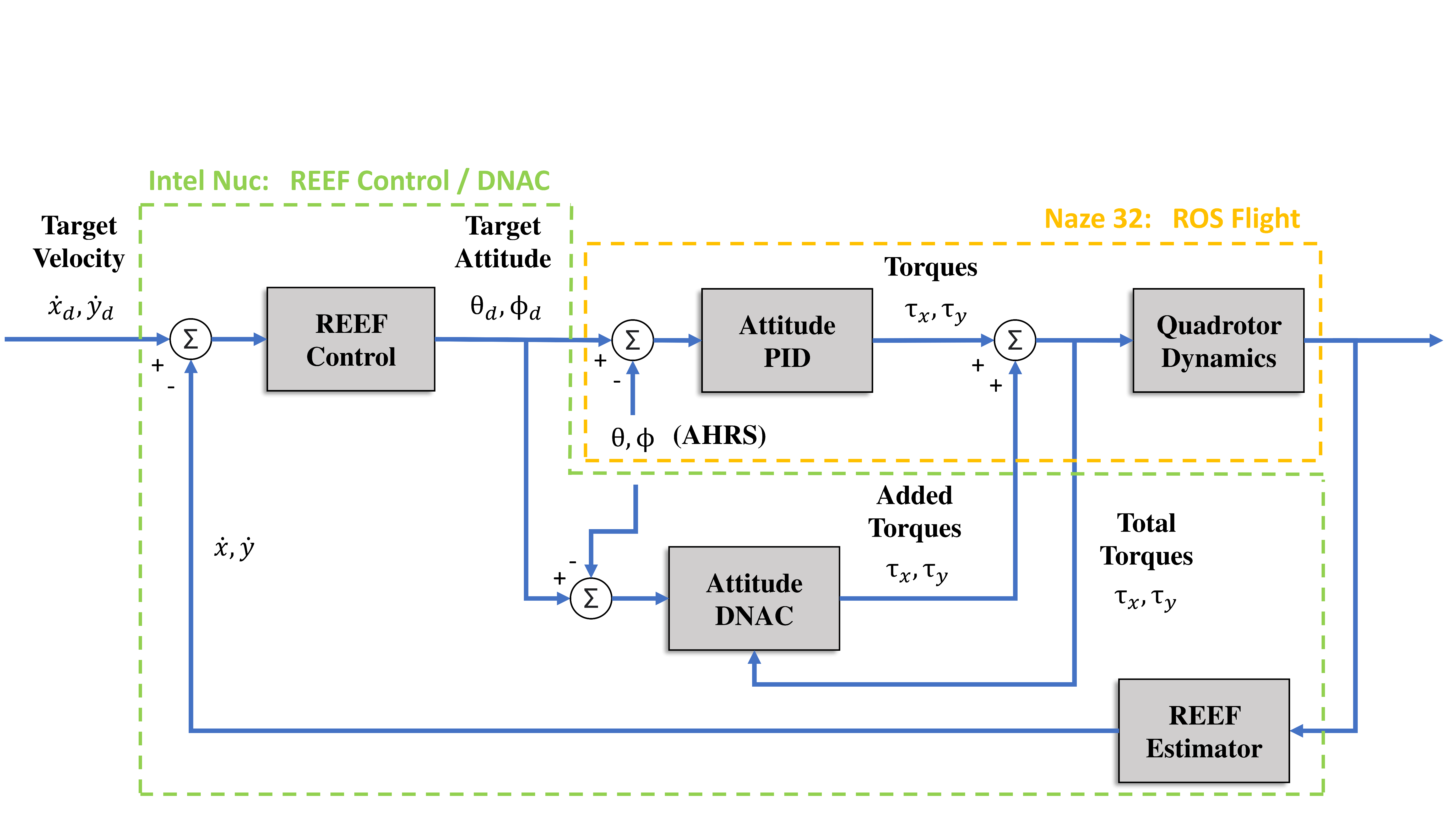}
    \caption{\footnotesize DNAC Augmented Attitude Control System - DNAC acts in parallel to the ROSflight attitude PID using ROSflight estimated attitudes and target attitudes produced by the REEF Control PID. REEF Control produces these target attitudes using a PID loop based on REEF Esimator velocities and target velocities. The control system is flexible to different target velocity generators, and in theses experiments, the target velocities are generated by a PID loop on position tracking error (not shown).}
    \label{dnacSimplified}
\end{figure*}

The experimental setup uses ROSflight \cite{rosFlight} - an autopilot firmware that is used to control the quadrotor. This specific firmware is useful because it has support for multiple Flight Control Units (FCU's) as well as a software-in-the-loop (SIL) simulation environment known as Gazebo. ROSflight provides a standard PID attitude control scheme which is naive to many of the linear model-based control issues mentioned in Section \ref{intro}. The ROSflight attitude PID scheme is improved upon via a parallel DNAC implementation which does not so much as replace the attitude level PID, but rather augments it. This architecture is preferred as it provides a baseline controller to facilitate stable control of the quadrotor in nominal environments, while allowing DNAC to assist when uncertain disturbances inject themselves into the system. This augmentation architecture has been used in \cite{anderson2023reef} for a similar MRAC implementation. Moreover, this augmentation scheme is easily reproducible at the position and velocity levels, allowing for multiple DNAC's to be run concurrently. Section \ref{ss:Augmentation} dives further into the aforementioned augmented ACS.\\

\subsection{Baseline System Architecture} \label{ss:system architecture}
The flight software in this work is run onboard a custom built quadrotor capable of carrying several sensors and powerful computational devices. The ROSFlight autopilot runs on an STM32 based Naze32 board and handles PID based attitude control of the drone; subsequent controllers are required to control position and velocity. The REEF Estimator/Controller package \cite{ramos2019reef} is a simplified velocity estimator and controller used to estimate the body-level velocity of the quadrotor by using data from a combination of sensors -- namely, an RGB-D camera, motion capture system, sonar and an IMU. These velocity estimates are fed into the REEF Control module which is implemented using a PID controller to output attitude and thrust commands. All of this processing and computation occurs on an Intel NUC companion computer which is also mounted onboard the quadrotor and serially linked to the FCU. The serial delay in communication is negligible in this setup.

Thrust and attitude commands are then fed into the ROSFlight attitude level PID controller, where they are compared against the onboard Attitude Heading Reference System (AHRS) estimates which are generated via a complementary filter \cite{Mahony}. Finally, the attitude level PID produces torque commands which are converted to motor PWM signals through a motor mixing algorithm. The reference signal for the REEF Controller comes from a position trajectory generator node. It is through this process that the REEF Estimator and Controller aid in commanding the drone to any desired position. A block diagram of this entire control system (with DNAC augmentation) is presented in  Figure \ref{dnacSimplified}.

\subsection{Augmented DNAC Attitude Control System} \label{ss:Augmentation}
The control configuration described in Subsection \ref{ss:system architecture} is typical for a quadrotor's flight control design, however, this configuration results in an inherent inability for the drone to adapt to changes in the model (i.e., a sudden gust of wind or a propeller is damaged). Moreover, there are also limitations to the range of attitude trajectories the drone can reliably track because of linear model assumptions that were made when the controller was designed. To improve the adaptability of the controller, DNAC can be used in parallel with the stable ROSflight attitude PID controller running onboard the FCU. Running neural networks on microcontrollers is difficult, thus DNAC is run onboard the companion computer. From there, DNAC operates on the same attitude commands and AHRS estimates as the ROSflight PID controller. It is in this way that the two separate controllers ensure they are operating on the same tracking error terms. Next, DNAC computes \textit{added torque} commands via control law (\ref{controlLaw1}), which are then sent to the FCU and added to the output of the ROSflight PID block. Finally, this is fed back into DNAC as a \textit{total torque} command which is necessary for learning the correct dynamics. Figure \ref{dnacSimplified} provides a visual of this implementation. 

The reasoning behind augmenting the current ROSflight autopilot and not replacing it is two-fold: 1) It is typically safer to restrict the amount of contribution an ML based controller can have on a system in case something goes wrong. If DNAC becomes dis-functional for any reason, the augmentation scheme can be severed and the stable PID attitude controller takes over 100\% of the flight control. With this augmentation scheme, it is up to the user how much they wish for DNAC to contribute to the ACS. 2) This augmentation design adds to the simplicity of implementation for other UAS's where stable flight controllers already exist and the user only wishes to see if performance may be improved without removing a large chunk of the baseline performing software.

Without loss of generality, the experiments run in the next section will focus on the roll and pitch of a quadrotor to demonstrate DNAC's ability to adaptively trim out a quadrotor under heavy wind and mass disturbances. Yaw is left out of the augmentation as it does not yield significant benefit because the applied disturbances tend to effect roll and pitch more-so.

\section{RESULTS}\label{results}
This section covers two experiments, where the purpose is to fly the quadrotor using various attitude control schemes under highly nonlinear dynamic uncertainties, and compare the results for each controller. Subsection \ref{ss:controlConfig} begins with DNAC's control configuration; this includes gain values, neural network architecture, and loss function choice. PID, MRAC, and DMRAC control configurations are discussed less extensively. Subsections \ref{ss:experiment1} and \ref{ss:experiment2} then cover experimental results and analyze the performance of each attitude controller. A standard ROSflight PID scheme is the baseline controller to which each additional controller augments. 

DNAC runs in parallel to this PID as previously described in Figure \ref{dnacSimplified}. Additionally, MRAC and DMRAC are implemented in a similar fashion, where the outputs of their respective control laws augment that of the original ROSflight PID. These four control schemes (PID, PID+MRAC, PID+DMRAC, PID+DNAC) are flown in a highly uncertain environment where disturbances include a crosswind, wall effect, hanging mass, sloshing effect from a partially filled water bottle. The attitude RMSE is compared for each controller, and position plots are presented as a visual aid. Note that the original ROSflight PID is capable of flying each of the trajectories in the experiments below given no disturbances are present. Moreover, each augmentation control scheme for all four controllers have been tested to work while operating under no model disturbances as well.

\subsection{Controller Configuration}\label{ss:controlConfig}
The original ROSflight PID scheme has been previously covered in \cite{rosFlight}. MRAC and DMRAC control schemes follow closely to the configurations found in \cite{girishDMRAC2}. The DNAC control configuration is as follows: Beginning with a definition of the state and input, define $x=[\phi , \theta]^\top$, and $u=[\tau_\phi, \tau_\theta]^\top$. From the system given in (\ref{dyn}), it follows that the input to the DNN estimate of the model uncertainty, $\widehat{f}(t,x)$, is the current roll and pitch of the drone. The setup of control law (\ref{controlLaw1}) begins with the assumption that both torque commands will actuate the vehicle with the same amount of control effectiveness. As such, a simple starting point for the control effectiveness matrix is $\widehat{g}(x(t))=I_2$. However, in practice $\widehat{g}(x(t))$ can be exploited as another tuning parameter since $\widehat{g}$ is inversley proportional to the outputted control effort. This relation allows for using $\widehat{g}$ as a method to reduce the overall effectiveness of $u$. In the following experiments,  $\widehat{g}=100\cdot I_2.$ While this approximation is likely inaccurate, the DNN will compensate for it in the model uncertainty term, $\widehat{f}(t,x)$. This model uncertainty is approximated via update law (\ref{fHat}), which uses a DNN to adjust the inner layer weights in batches and a nonlinear adaptive law (\ref{controlLaw2}) to adjust the outer layer weights in real-time. The real-time learning gain is set as $\Gamma_W=10$. This term effects the outer layer weight estimates similar to the step size in gradient descent learning strategies. Finally, the proportional control gains are set as $K=10$ and $K_s=0.001$. In practice, $K_s$ is kept small. 

The slower timescale learning is accomplished by utilizing PyTorch with a network structure $\widehat{\Phi}: \mathbb{R}^2 \rightarrow \mathbb{R}^8$ which is decomposed for each layer as $\widehat{\Phi}(x) = ( \widehat{\phi}_3 \circ \widehat{\phi}_2 \circ \widehat{\phi}_1 ) (x) $ where $\widehat{\phi}_1: \mathbb{R}^2 \rightarrow \mathbb{R}^3, \widehat{\phi}_2: \mathbb{R}^3 \rightarrow \mathbb{R}^4, \widehat{\phi}_3: \mathbb{R}^4 \rightarrow \mathbb{R}^8$ and $\widehat{\phi}_{l} (x) = \sigma_{l} ( \widehat{W}_{l}^\top x + \widehat{b}_{l} )$. The associated activation functions after each linear layer are $\sigma_1$ = $\text{tanh}$, $\sigma_2$ = $\text{log-sigmoid}$, and $\sigma_3$ = $\text{tanh}$. The network is finally transformed by the outer layer weights, $\widehat{W}$, to map the output back into the $\mathbb{R}^2$ space as $\widehat{W}^\top \sigma(\widehat{\Phi}(x))$.

The replay memory of size $M=100$ is split evenly into smaller batches of size $S_b=20$, and the 5 segments are shuffled randomly. The state $x$ (roll and pitch) is extracted from these mini batch segments, and are passed through the network $N_e=5$ (number of epochs) times each during training cycles. This training results in an $\widehat{f}(x_k)$ term which is used alongside batch data $\widehat{g}(x_k)u_k$ to estimate $\dot{\widehat{x}}_k$ (the roll and pitch rates of the drone) via (\ref{xPred}). Finally, an optimization step is taken to minimize the smooth L1 loss between the network’s approximation of $\dot{\widehat{x}}_k$ and the measured $\dot{x}_k$, which is present in the replay buffer. A summary of these controller parameters is provided in Table \ref{controlParameter}.

\begin{cmr}
\footnotesize
\begin{center}
\begin{tabular}{| p{2cm} p{2cm} p{3.4cm} |}
\hline
Parameter & Value & Description\\
\hline
\rowcolor{grey}
 \text{$K$} & 10 & Control Gain\\
 \text{$K_{s}$} & 0.001 & Control Gain\\
 \rowcolor{grey}
 \text{$\widehat{g}$} & $100\cdot I_{2}$ & Model Assumption \\  
 \text{$\Gamma_{W}$} & 10 & Online Learning Gain\\
 \rowcolor{grey}
 \text{$M$} & 100 & Batch Memory Size \\
 \text{$S_b$} & 20 & Training Batch Size\\
 \rowcolor{grey}
 \text{$N_e$} & 5 & Number of Epochs\\
 \hline
\end{tabular}\captionof{table}{\footnotesize DNAC Controller parameters related to control law (\ref{controlLaw1}), adaptive update law (\ref{controlLaw2}), and DNN architecture.}\label{controlParameter}
\end{center}
\end{cmr}

\begin{figure*}[h]
    \centering 
    \includegraphics[width=6.5in]{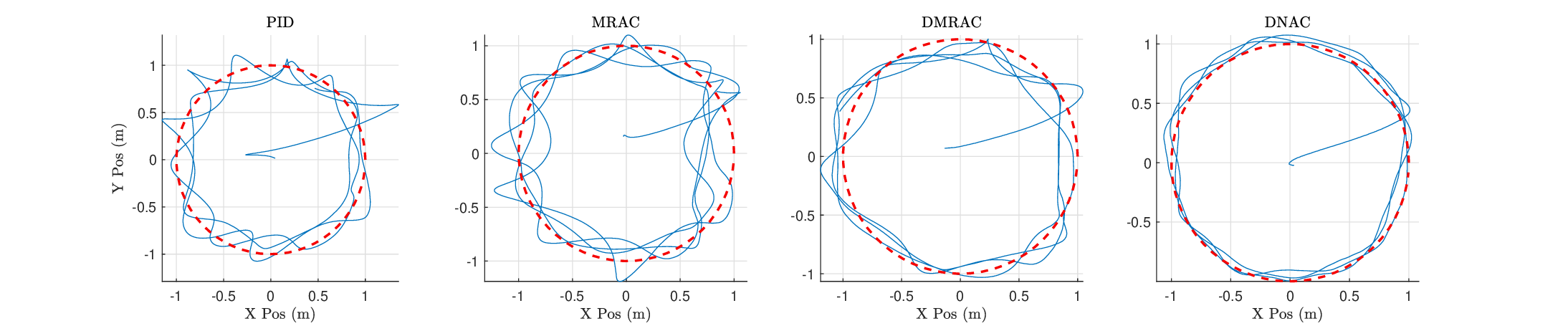}
    \includegraphics[width=6.5in]{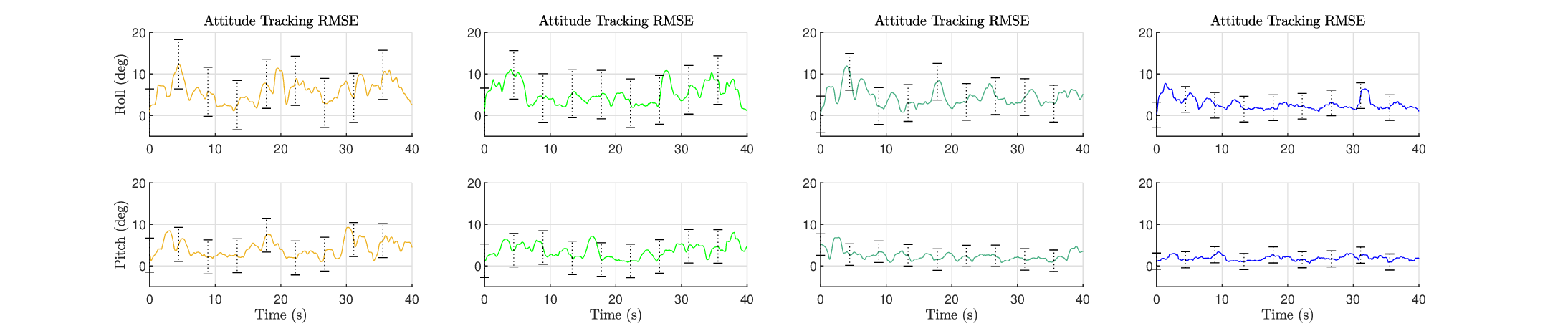}
    \caption{\footnotesize Top row: Quadrotor desired trajectory (red dashes) and true trajectory (blue) results for the hardware experiment. Model disturbance rejection is visualized by the circularity of the quadrotor’s true trajectory. Bottom row: Quadrotor moving RMS attitude tracking error (solid line) and standard deviation (vertical bars). From left to right: (i) baseline PID performance (ii) baseline PID with MRAC augmentation (iii) baseline PID with DMRAC augmentation (iv) baseline PID with DNAC augmentation.}
    \label{fig:temp_circles}
\end{figure*}

Note that the DNN in the following experiments is initialized with random weights, and the size of the network is relatively small when compared to other DNN applications; despite these facts, experimental results prove this method to be effective at learning the system dynamics in real time with no prior-training. This is possible because the Lyapunov-based outer-layer weight updates allow the system to learn the most important features of its dynamics in real-time. This is opposed to traditional DNN learning approaches which are based solely on batch updates. Additionally, the inner-layer weights are trained on a loss between the measured time derivative of the state (angular roll and pitch rate in this case) and the DNN based approximate of the time derivative of the state, while the outer-layer weights are learned based on current inner-layer features as well as the current trajectory tracking error. These two reasons -- real-time updates to the model uncertainty term and a learning strategy based on the state derivative loss as well as on trajectory tracking error -- primarily serve as the basis for DNAC's strong ability to learn new dynamics quickly. Importantly, because a measurement (or estimate) of the state derivative with respect to time is required, controlling velocity or acceleration may prove more difficult, as it requires an accurate estimate of acceleration or jerk respectively.

\subsection{Experiment 1}\label{ss:experiment1}
Using the quadrotor control setup outlined in Section \ref{controlPlatform}, the quadrotor is flown in a 1 m radius circular trajectory with a water bottle filled to 125 mL hanging from a string attached at the base of the motor (located 0.25 m from the center of mass) in the positive pitch – positive roll quadrant. A large fan (36 in diameter) is then placed 2 m from the center of the circle trajectory, and creates an 18 mph crosswind. Recall that the quadrotor is tracking velocity commands generated via cascading PID loops. These velocity PID controllers generate attitude references which the drone's augmented ACS attempts to track. As such, positional tracking is not a direct measurement of each controller’s performance; however, it is still an indirect indicator to evaluate performance. 

The circularity of the quadrotor's trajectory (Figure \ref{fig:temp_circles}) is a visual evaluation of each control scheme's (PID, MRAC, DMRAC, DNAC) robustness to disturbances. Under the conditions, each augmentation controller improves upon baseline PID performance (as quantified in Table \ref{trackingErrorNorm}); this is evident by the removal of the cusps in the baseline PID trajectory which

\begin{cmr}
\footnotesize
\begin{center}
\begin{tabular}{| p{1.5cm} p{1.8cm} p{1.8cm} p{2cm} |}
\hline
Controller & $\lVert \theta,\phi \rVert_2 (^\circ)$ & $\lVert x,y \rVert_2 (\text{cm})$ & $\lVert \dot{x},\dot{y} \rVert_2 (\text{cm/s})$\\
\hline
\rowcolor{grey}
 \text{PID} & 6.88 & 30.0 & 40.2\\
 \text{MRAC} & 6.13 & 26.7 & 32.0\\
 \rowcolor{grey}
 \text{DMRAC} & 4.93 & 24.4 & 29.4\\  
 \text{DNAC} & 3.06 & 16.6 & 14.9\\
 \hline
\end{tabular}
\captionof{table}{\footnotesize Average tracking error $L_2$ norms for flight data taken while the drone tracks a circular trajectory in Experiment 1.}\label{trackingErrorNorm}
\end{center}
\end{cmr}

\begin{cmr}
\footnotesize
\begin{center}
\begin{tabular}{| p{2cm} p{1.5cm} p{1.5cm}|}
\hline
Controller & $\sigma_\phi (^\circ)$ & $\sigma_\theta (^\circ)$\\
\hline
\rowcolor{grey}
 \text{PID} & 5.92 & 4.09 \\
 \text{MRAC} & 5.85 & 4.02\\
 \rowcolor{grey}
 \text{DMRAC} & 4.42 & 2.59\\  
 \text{DNAC} & 3.08 & 1.93\\
 \hline
\end{tabular}
\captionof{table}{\footnotesize Standard deviation of the attitude tracking error for each controller configuration in Experiment 1.}\label{trackingErrorStdDev}
\end{center}
\end{cmr}

\noindent is most obvious when comparing PID and MRAC performances. The deep controller schemes go further and remove the PID’s oscillatory behavior with varying levels of success. The DNAC scheme not only rejects oscillatory behavior better than the DMRAC scheme, but the DNAC scheme also reduces the position tracking error norm to 16.6 cm which is a 32\% decrease from DMRAC, a 37.8\% decrease from MRAC, and a 44.7\% decrease from PID. Furthermore, DNAC reduces the velocity tracking error norm to 14.9 cm/s which is a 49.3\% decrease from DMRAC, 53.4\% decrease from MRAC, and a 63\% decrease from PID. 

While the visual improvements from each position plot are insightful, a more relevant performance metric is to measure the error between the quadrotor's attitude trajectory and the reference attitude trajectory. The attitude error norm results for each controller show that the DNAC implementation improves upon DMRAC performance by an error decrease of 37.9\%, MRAC performance by an error decrease of 50.8\%, and baseline PID performance by an error decrease of 55.5\%. Furthermore, DNAC yields a lower standard deviation on attitude tracking error than the other controllers. The standard deviation of each controller (indicated by black bars in Figure \ref{fig:temp_circles}) is summarized in Table \ref{trackingErrorStdDev}.

\subsection{Experiment II}\label{ss:experiment2}
In the second experiment, a new desired trajectory is designed and another model disturbance is introduced. The quadrotor now flies in a four-petal rose pattern described by $r = a \sin(2\theta)$, where $a=2.8$. The same model disturbances from Experiment 1 are present, as well as an additional wall effect disturbance; that is, two L-shaped wall's surround the two left-most rose-petals in the trajectory in order induce turbulent airflow when the quadrotor approaches, while the fan largely effects the two right petals with an 18 mph source stream of wind and the filled water bottle inputs stochastic torques into the system at all times. The purpose of this experiment is to present DNAC's ability to handle learning sudden and brief disturbances; the quadrotor will attempt to track a trajectory for which the box fan covers a partial span of the flight path, and the walls which the quadrotor must navigate near span a separate part of the flight path. These conditions are unlike experiment 1, where the 18mph wind gusts were relatively constant throughout the entire flight path, and the flight path itself was much smaller. Moreover, because the flight regime in experiment 2 is larger, the quadrotor will be commanded to deviate further from the linearized hover state. This leads to the deduction that DNAC, a nonlinear controller, should outperform DMRAC, a linear controller, in a more obvious manner.

\begin{figure}[h]
    \centering 
    \includegraphics[width=3.5in]{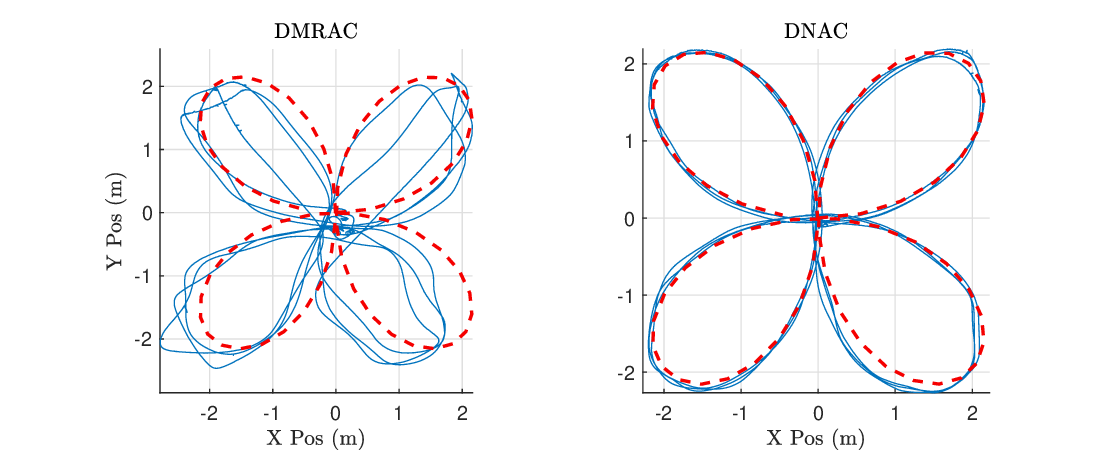}
    \includegraphics[width=3.5in]{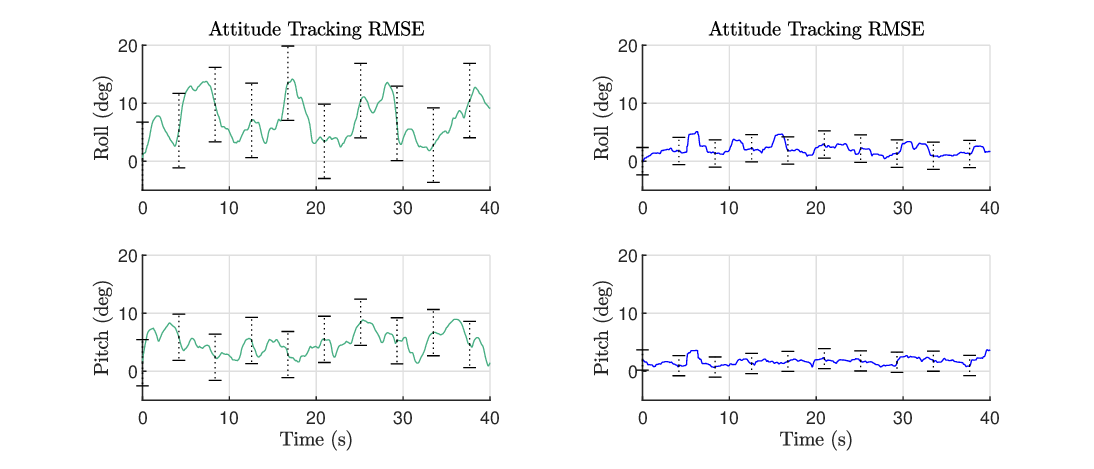}
    \caption{\footnotesize Top left: DMRAC position tracking. Top right: DNAC position tracking. Bottom left: DMRAC attitude tracking RMSE and standard deviation. Bottom right: DNAC attitude tracking RMSE and standard deviation. In the position tracking plots, the wind disturbance field originates in the positive-$X$ negative-$Y$ quadrant blowing upward, while the two L-shaped walls are located in the corners of the trajectory in the negative-$X$ quadrant.}
    \label{fig:four-leaf}
\end{figure}

Experiment 1 and \cite{girishDMRAC2} both conclude DMRAC outperforms PID and MRAC in a quadrotor autopilot application. For this reason, and for brevity, these control schemes are left out of the experiment 2 analysis. Similar to Experiment 1, DNAC is initialized with random weights and has not been further tuned. This is a key point in the analysis of DNAC's ability to adapt to a variety of situations with little to no software/algorithm change. The position and attitude tracking RMSE results of the four-petal rose experiment are provided in Figure \ref{fig:four-leaf}.


With this new flight path, DMRAC has a difficult time handling the variety of sudden disturbances, and performs worse than when the disturbances remained more constant over time. This is not the case for DNAC as it learns to reject each of the disturbances quickly while moving across them. From Table \ref{trackingErrorNormExp2}, it is observed that DNAC tracks desired attitude with greater accuracy than both DRMAC from experiment 2, as well as DNAC from experiment 1; specifically, DNAC tracks the four-petal rose pattern with an attitude error L2 norm of $2.10^\circ$. This is a $6.21^\circ$ improvement over DMRAC. Moreover, the standard deviation of the attitude error (provided in Table \ref{trackingErrorStdDevExp2}) for DNAC shows substantial improvement over DMRAC. With respect to experiment 1, DNAC exhibits a $0.96^\circ$ decrease in attitude tracking RSME as well as a decrease in standard deviation. These performance boosts may in part be due to the additional system modes being excited along the larger and more complex flight path.

\begin{cmr}
\footnotesize
\begin{center}
\begin{tabular}{| p{1.5cm} p{1.8cm} p{1.8cm} p{2cm} |}
\hline
Controller & $\lVert \theta,\phi \rVert_2 (^\circ)$ & $\lVert x,y \rVert_2 (\text{cm})$ & $\lVert \dot{x},\dot{y} \rVert_2 (\text{cm/s})$\\
\hline
\rowcolor{grey}
 \text{DMRAC} & 8.31 & 70.0 & 49.2\\  
 \text{DNAC} & 2.10 & 47.0 & 19.1\\
 \hline
\end{tabular}
\captionof{table}{\footnotesize Average tracking error $L_2$ norms for flight data taken while the drone tracks a four-petal rose trajectory in Experiment 2.}\label{trackingErrorNormExp2}
\end{center}
\end{cmr}

\begin{cmr}
\footnotesize
\begin{center}
\begin{tabular}{| p{2cm} p{1.5cm} p{1.5cm}|}
\hline
Controller & $\sigma_\phi (^\circ)$ & $\sigma_\theta (^\circ)$\\
\hline
 \rowcolor{grey}
 \text{DMRAC} & 6.41 & 3.98\\  
 \text{DNAC} & 2.35 & 1.73\\
 \hline
\end{tabular}
\captionof{table}{\footnotesize Standard deviation of the attitude tracking error for each controller configuration in Experiment 2.}\label{trackingErrorStdDevExp2}
\end{center}
\end{cmr}

It should be noted here that trajectory tracking performance for DNAC in both experiments is improving over time. As the inner layer features are trained on a slower timescale, tracking bias is removed on each cycle of the flight path. This presents itself in the position tracking plots where one see's (forward in time) a more tightly followed reference trajectory.

\section{CONCLUSION}
\begin{figure}[h]
    \centering
    \includegraphics[width=2.5in,height=2in]{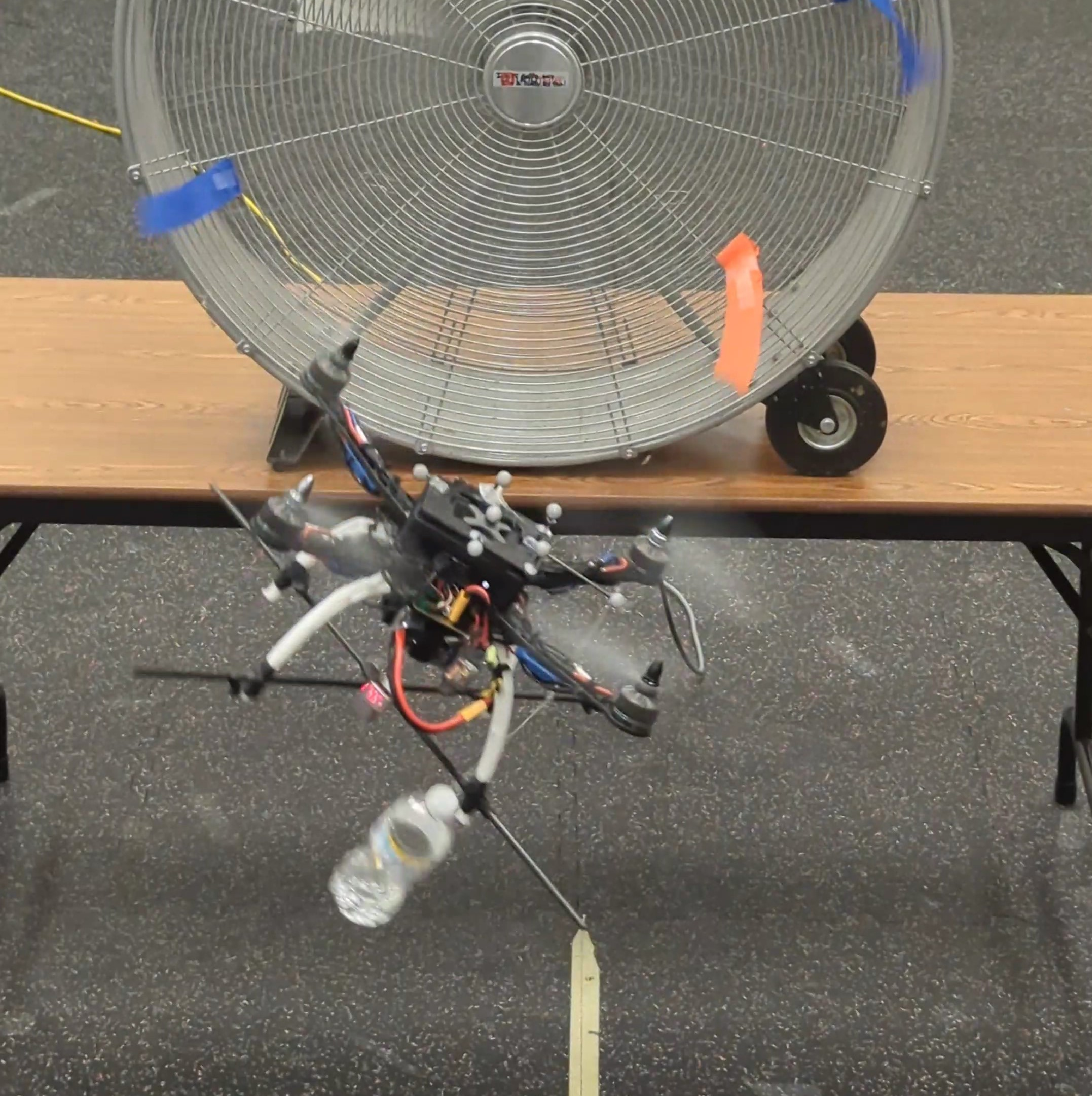}
    \caption{Quadrotor demonstrating the magnitude of the effect of the sloshing hanging mass and wind disturbances. The quadrotor is shown 1 m from the fan.}
    \label{fig:quadrotorSetup}
\end{figure}

Linear model-based rigid body controllers demonstrate drawbacks for aerospace systems which can be remedied through nonlinear adaptive control techniques. Background was provided for recent advancements in the literature which introduce Deep Neural Network based adaptive control techniques with stability guarantees -- namely, Deep Model Reference Adaptive Control (DMRAC) \& Deep Nonlinear Adaptive Control (DNAC). The control laws governing DNAC were then presented, and an augmented attitude control system was devised for a quadrotor. Flight experiments were run in which a variety of attitude controller performances were compared while operating in a highly uncertain flight environment (see Figure \ref{fig:quadrotorSetup}). These experiments show DNAC outperforming other controllers with an ability to easily learn and reject undesirable model disturbances without a reliance on prior training data or prior identification of some set of system parameters. This paper 1) demonstrated the design of an ML based controller capable of leveraging the powerful model estimation ability of a DNN whilst maintaining stability guarantee's necessary to control a safety-critical system with no reliance on prior training data, and 2) offered an easily configurable and tunable implementation of a deep nonlinear adaptive controller which can run in parallel to existing control structures and improve performance even at the lowest levels of the feedback loop. DNAC achieves this performance while only requiring basic knowledge about the system and simple tuning processes, showing the potential of the DNAC architecture to improve performance in other safety critical systems operating in highly uncertain environments.\\\\

\bibliographystyle{IEEEtran}
\bibliography{reef,dnns}


\end{document}